\documentstyle{article}
\begin{document}
\centerline{\large\bf The Hydra Consortium}
\centerline{\bf Purveyors of fine N-body algorithms and simulation data}
\smallskip
\centerline{\tt http://coho.astro.uwo.ca/pub/consort.html}
\centerline{\tt http://star-www.maps.susx.ac.uk/$\sim$frp/consort.html}
\bigskip

The Hydra Consortium exists to provide public domain software and data
for N-body, hydrodynamical simulations.

We value your feeback, whether positive or critical.  In particular,
if you make use of any of our products {\bf please register}
by email to the appropriate author.
\bigskip

\leftline{\bf The Hydra Consortium:}
\begin{itemize}
\item Hugh Couchman\newline
{\tt http://coho.astro.uwo.ca/pub/couchman/home.html}\newline
{\tt mailto:couchman@coho.astro.uwo.ca}
\item Frazer Pearce\newline
{\tt http://star-www.maps.susx.ac.uk/people/frp.html}\newline
{\tt mailto:frazerp@central.susx.ac.uk}
\item Peter Thomas\newline
{\tt http://star-www.maps.susx.ac.uk/people/pat.html}\newline
{\tt mailto:p.a.thomas@sussex.ac.uk}
\end{itemize}
\bigskip
The authors would like to thank NATO for providing a Collaborative
Research Grant (CRG 920182) which facilitated our interaction.  The
facilities used to develop this code were provided by NSERC in Canada
and PPARC in the UK.  HMPC is supported by NSERC.  FRP is supported by
EPSRC as the PDRA for the Virgo Consortium.  This work
was carried out while PAT was holding a Nuffield Foundation Science
Research Fellowship.
\end{document}